\documentclass[epj]{webofc}
\usepackage[utf8]{inputenc}
\usepackage[varg]{txfonts}   
\usepackage{booktabs}
\usepackage{xcolor}
\usepackage{braket}
\usepackage{bbold}
\usepackage{caption}

\usepackage{multirow}
\definecolor{darkred}{rgb}{0.4,0.0,0.0}
\definecolor{darkgreen}{rgb}{0.0,0.4,0.0}
\definecolor{darkblue}{rgb}{0.0,0.0,0.4}
\usepackage[bookmarks,linktocpage,colorlinks,
    linkcolor = darkred,
    urlcolor  = darkblue,
    citecolor = darkgreen]{hyperref}
\usepackage{subfigure}
\wocname{EPJ Web of Conferences}
\woctitle{Lattice2017}
\begin{document}
%
\selectlanguage{english}
\title{
    Continuum extrapolation of quarkonium correlators at non-zero temperature
}
\author{%
\firstname{Heng-Tong} \lastname{Ding}\inst{1}\fnsep \and
\firstname{Olaf} \lastname{Kaczmarek}\inst{1,2} \and
\firstname{Anna-Lena} \lastname{Kruse}\inst{2} \and
\firstname{Hiroshi} \lastname{Ohno}\inst{3,4} \and
\firstname{Hauke}  \lastname{Sandmeyer}\inst{2}\fnsep\thanks{Speaker, 
\email{hsandmeyer@physik.uni-bielefeld.de}}
}
\institute{%
    Key Laboratory of Quark \& Lepton Physics (MOE) and Institute of Particle Physics,
    Central China Normal University, Wuhan 430079, China
    \and
    Fakultät für Physik, Universität Bielefeld, 33615 Bielefeld, Germany
    \and
    Physics Department, Brookhaven National Laboratory, Upton, NY 11973, USA
    \and
    Center for Computational Sciences, University of Tsukuba, Tsukuba, Ibaraki 305-8577, Japan
}
\abstract{%
    In the investigation of in-medium modifications of quarkonia and for determining heavy quark
diffusion coefficients, correlation functions play a crucial role. For the
first time we perform a continuum extrapolation of charmonium and bottomonium correlators in the vector 
channel based on non-perturbatively clover-improved Wilson fermions in quenched
lattice QCD. Calculations were done at 4 different lattice spacings, spatial extents between 96
and 192, aspect ratio from 1/6 to 1/2, for 5 temperatures between $T/T_c = 0.75$ and $T/T_c = 2.2$. We
interpolate between different quark masses to match to the same vector meson mass over different
lattice setups. Afterwards we extrapolate the renormalized correlators to the continuum. 
While we find a strong temperature dependence for charmonium, bottomonium states are only slightly
affected at higher temperatures.
}
\maketitle
\section{Introduction}\label{intro}
In the investigation of properties of the quark gluon plasma (QGP), which is created at heavy
ion collisions, quarkonium bound states are of great interest. Due to the color Debye screening, these
states are expected to melt at a certain dissolution temperature and, therefore, suppression of
$J/\psi$ yields at heavy ion collisions compared to p-p-collisions can be a signal for the
existence of a quark gluon plasma \cite{Matsui:1986dk}.

Indeed, $J/\psi$-suppression has been found at all three large colliders SPS, RHIC and LHC and
also $\Upsilon$ suppression has been reported at the LHC
\cite{Arnaldi:2009ph,Adare:2008qa,Abelev:2012rv,Aad:2010aa,Chatrchyan:2012np,
Chatrchyan:2012lxa}. Nevertheless, for a better theoretical interpretation of
the experiments, one needs more information about the in-medium modification of quarkonium 
bound states. 

In addition, measurements at RHIC and LHC revealed a non-zero elliptic flow for heavy
quarks \cite{Abelev:2013lca,Adare:2014rly,Vertesi:2014tfa}. This indicates a relaxation towards
thermal equilibrium and gives evidence to a collective motion of heavy quarks. For a better
understanding of these hydrodynamic effects, knowledge about the transport properties
of heavy quarks within the medium is needed. Especially, the heavy quark diffusion
coefficient $D$ is of interest, as it is related to the energy loss of a heavy quark in
the medium. So far leading order perturbative calculations are not in agreement with the
experimental data \cite{Moore:2004tg} and next-to-leading order calculations show that one can
expect large corrections by higher orders \cite{CaronHuot:2008uh}. Therefore, first principle
studies, such as lattice QCD, are necessary to understand the non-perturbative effects in these quantities.

Both, in medium modification as well as transport properties are encoded in the spectral
function of quarkonia. Unfortunately, the spectral function and the corresponding Euclidean correlator are related by an integral
and a lot of details of the spectral function are averaged away. Moreover, due to the limitation of the $\tau$-direction, only
few lattice points are available. Therefore, very fine lattices
as well as good statistics are needed to extract structures of the spectral function. So far,
there were several attempts using very fine quenched lattices, though no continuum extrapolation
has been realized \cite{Ding:2012iy,Ding:2012sp,Ohno:2014uga}. In \cite{Francis:2015daa} a 
continuum extrapolation for the heavy quark momentum diffusion coefficient based on a color
electric correlator has been carried out. However, this approach is limited to the heavy quark mass limit and as 
this analysis is not based on meson
correlators, in-medium properties of bound states cannot be extracted by this approach.

For the first time, we have carried out a continuum extrapolation for charmonium and 
bottomonium correlation functions in quenched lattice QCD. In this paper we will show the
details and steps that have been been used to perform the continuum extrapolation. A similar analysis for light
quarks can be found in \cite{Ding:2016hua,Ghiglieri:2016tvj}.

\section{Meson correlation functions and the spectral function}
The main object of interest in this study, the Euclidean mesonic correlation function, is
defined as
\begin{equation}
    G_{H}(\tau,\mathbf{x})=\braket{J_{H}(\tau,\mathbf{x})J_{H}^\dagger(0,\mathbf{0})}
\end{equation}
where the mesonic current is given by
\begin{equation}
    J_{H}(\tau,\mathbf{x})=\psi^\dagger(\tau,\mathbf{x})\Gamma_H\psi(\tau,\mathbf{x}).
\end{equation}
Here, $\psi$ are the fermion fields and $\Gamma_H=\gamma_5,\gamma_\mu,\mathbb{1},\gamma_5 \gamma_\mu$
corresponds to the meson channel we are looking at
(pseudo-scalar (P), vector (V), scalar (S), axial-vector (AV), respectively). In this study, we perform the continuum
extrapolation for the example of the vector channel, but, except for the renormalization, the process
works analogously for the other channels. For results for the pseudo-scalar channel, see \cite{Burnier:2017bod}.

The above correlator is projected to finite momentum using the lattice Fourier transformation,
\begin{equation}
    G_{V}(n,\mathbf{p})=\sum_\mathbf{x}G_{V}(\tau,\mathbf{\tilde{x}})\exp(i\mathbf{\tilde{p}}\mathbf{\tilde{x}})
\end{equation}
with $n=\tau$, $\mathbf{\tilde{x}}=(x,y,z)$ and $\mathbf{\tilde{p}}=(p_x,p_y,p_z)$ for temporal correlators or
$n=z$, $\mathbf{\tilde{x}}=(x,y,\tau)$ and $\mathbf{\tilde{p}}=(p_x,p_y,p_\tau)$ for spatial correlators, respectively.
In the following we restrict to $\mathbf{\tilde{p}}=\mathbf{0}$ and, except for the mass determination,
to temporal correlators.

The temporal correlator is connected to the spectral function via
\begin{equation}
G_V(\tau)=\int_0^\infty\rho_V(\omega)K(\omega,\tau)\mathrm{d}\omega,\quad
K(\omega,\tau)=\frac{\cosh(\omega(\tau-1/2T))}{\sinh(\omega/2T)}.
\end{equation}
The spectral function contains all information on the medium properties of mesons. Whereas many
information, like dissociation temperatures, can be extracted from the deformation of the spectral
functions at higher temperature, the diffusion constant $D$ can be directly extracted from the 
vector spectral function:
\begin{equation}
    D=\frac{\pi}{3\chi_q}\lim_{\omega\rightarrow 0}\sum_{i=1}^3\rho_{ii}(\omega,T)/\omega.
\end{equation}
Thereby, the index at the spectral function stems from the $\mu$-index from $\Gamma_V=\gamma_\mu$ and
$\chi_q$ is the quark number susceptibility given by
\begin{equation}
    \chi_q={G_V}_{00}(\tau)/T.
\end{equation}
While this is in principle constant in $\tau$ due to current conservation,
lattice cut-off effects are noticeable for short $\tau$-ranges. Therefore we extract $\chi_q$ from 
${G_V}_{00}(\tau T=0.5)$.

Since the quark number susceptibility is affected by the same renormalization as ${G_V}$,
it is used to circumvent the uncertainties of the renormalization constants using
renormalization independent ratios by dividing all correlators by the quark number susceptibility
at $T'=2.25T_c$.

In order to remove the exponential drop of the correlator we normalize 
all correlators by the free correlator which is given by
\begin{align*}
    G_{\mathrm{free}}(\tau)/T^3 &= \int_{2m_q}^\infty \frac{\rho_\mathrm{free}(\omega, m_q)}{T^3} K(\omega,\tau )d\omega,\\
    \rho_\mathrm{free}(\omega, m_q) &= \frac{3}{16\pi^2}\omega^2\tanh\left(\frac{\omega}{4T}\right)\sqrt{1-\left(\frac{2m_q}{\omega}\right)^2}\left(4+2\left(\frac{2m_q}{\omega}\right)^2\right).
    \label{eq:Gfree}
\end{align*}
The choice of $m_q$ is arbitrary in this case, as this normalization is only temporary and will be removed
after the final continuum extrapolation. In the following we use $m_q=1.5$GeV for charmonium and
$m_q=5$GeV for bottomonium.

\section{Lattice setup}

\begin{figure}
    \centering
    \includegraphics[width=0.49\textwidth]{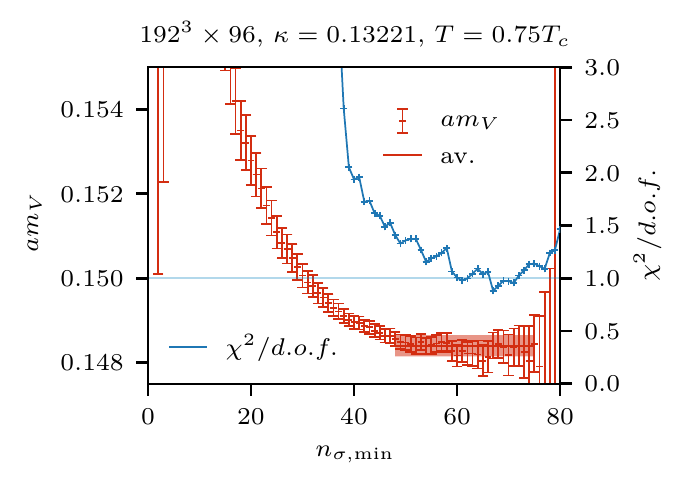}
    \includegraphics[width=0.49\textwidth]{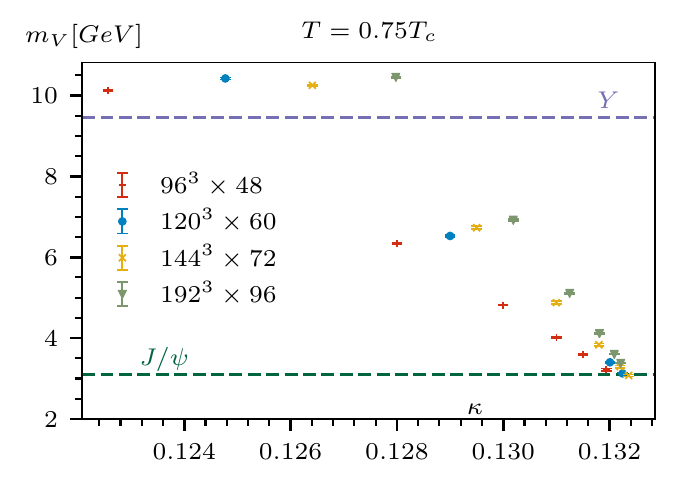}
    \caption{Left: Screening mass plateau for different fit intervals of size $[n_{\sigma,\mathrm{min.}}, n_{\sigma, \mathrm{max}}]$
        for the finest lattice. $n_{\sigma, \mathrm{max}}$ is fixed to 88 for this lattice size.
    Right: $\kappa$-values and the corresponding vector meson mass that have been measured.}
    \label{fig:masses}
\end{figure}
The correlators were measured on quenched configurations generated with the standard Wilson gauge action on large isotropic lattices. All configurations
are separated by 500 sweeps which consist of one heatbath and four overrelaxation steps.
Details about configurations can be found in table \ref{tab:lattice_setup}. The spatial extent $N_\sigma$ varies from 96 to 192
and the corresponding temporal extent has been chosen to match the temperatures $T/T_c=0.75,1.1,1.3,1.5,2.25$.
The volumes have been chosen in such a way that we get the same fixed aspect ratios, which results
in the same physical volume ($8.43\mathrm{GeV}^{-1}$) for all lattices.
The lattice scale has been determined by the Sommer parameter $r_0$ from \cite{Francis:2015lha} 
with updated coefficients to include a wider $\beta$-range \cite{Burnier:2017bod}.

The correlators have been calculated using non-perturbatively clover-improved Wilson fermions \cite{Sheikholeslami:1985ij}.
We computed the correlators for six different $\kappa$-values for all lattices except for the $N_\sigma=120$
lattice, where we used four different $\kappa$-values. The $\kappa$-values have been roughly distributed
between bottomonium and charmonium using the vector screening masses from the spatial correlator 
in the deeply confined phase at $0.75T_c$ for tuning.
The screening masses have been determined by averaging over the masses from correlated two state fits for different fit intervals.
As a fit Ansatz we use
\begin{align}
    G(n_\tau) = A_1\cosh(m_1(n_\tau - N_\tau/2))
    + A_2\cosh(m_2(n_\tau - N_\tau/2)).
\end{align}

For the fit intervals, we fixed the upper limit to $N_\sigma/2-N_\sigma/24$ and gradually increased the lower limit.
Afterwards we averaged over all fit results that lied within a plateau of the screening mass and
in the corresponding $\chi^2/d.o.f.$ of the fit (See figure \ref{fig:masses}). Due to the high costs
of calculating the correlators, the quark mass tuning is not optimal.
This will be corrected by a quark mass interpolation later on.

For the finest lattice, we used four additional quark sources for the two $\kappa$-values closest to bottomonium
and charmonium to increase the statistics at these important data points.

\begin{table}[t]
    \centering
    \footnotesize{
        \begin{tabular}{ccccccc}
\hline 
$\beta$ & $a${[}fm{]}($a^{-1}${[}GeV{]}) & $\kappa$-range & $N_{\sigma}$ & $N_{\tau}$ & $T/T_{c}$ & \#meas.\tabularnewline
\hline 
\hline 
\multirow{5}{*}{7.192} & \multirow{5}{*}{0.018 (11.19)} & \multirow{5}{*}{0.12257-0.13194} & \multirow{5}{*}{96} & 48  & 0.75  & 237\tabularnewline
 &  &  &  & 32  & 1.1  & 476\tabularnewline
 &  &  &  & 28  & 1.3  & 336\tabularnewline
 &  &  &  & 24  & 1.5  & 336\tabularnewline
 &  &  &  & 16 & 2.25 & 237\tabularnewline
\hline 
\multirow{4}{*}{7.394} & \multirow{4}{*}{0.014 (14.24)} & \multirow{4}{*}{0.124772-0.132245} & \multirow{4}{*}{120} & 60  & 0.75  & 171\tabularnewline
 &  &  &  & 40  & 1.1  & 141\tabularnewline
 &  &  &  & 30  & 1.5  & 247\tabularnewline
 &  &  &  & 20 & 2.25 & 226\tabularnewline
\hline 
\multirow{5}{*}{7.544} & \multirow{5}{*}{ 0.012 (17.01)} & \multirow{5}{*}{0.12641-0.13236} & \multirow{5}{*}{144} & 72  & 0.75  & 221\tabularnewline
 &  &  &  & 48  & 1.1  & 462 \tabularnewline
 &  &  &  & 42  & 1.3  & 660\tabularnewline
 &  &  &  & 36  & 1.5  & 288\tabularnewline
 &  &  &  & 24 & 2.25 & 237\tabularnewline
\hline 
\multirow{5}{*}{7.793} & \multirow{5}{*}{0.009 (22.78)} & \multirow{5}{*}{0.12798-0.13221} & \multirow{5}{*}{192} & 96  & 0.75  & 224\tabularnewline
 &  &  &  & 64  & 1.1  & 291\tabularnewline
 &  &  &  & 56  & 1.3  & 291\tabularnewline
 &  &  &  & 48  & 1.5  & 348\tabularnewline
 &  &  &  & 32 & 2.25 & 235\tabularnewline
\hline 
\end{tabular}

    }
    \caption{
        Lattice setup used to perform the continuum extrapolation.
        For the finest lattice, we used five quark sources for the two kappa values closest to bottomonium and charmonium
    \label{tab:lattice_setup}}
\end{table}

\begin{figure}
    \centering
    \includegraphics{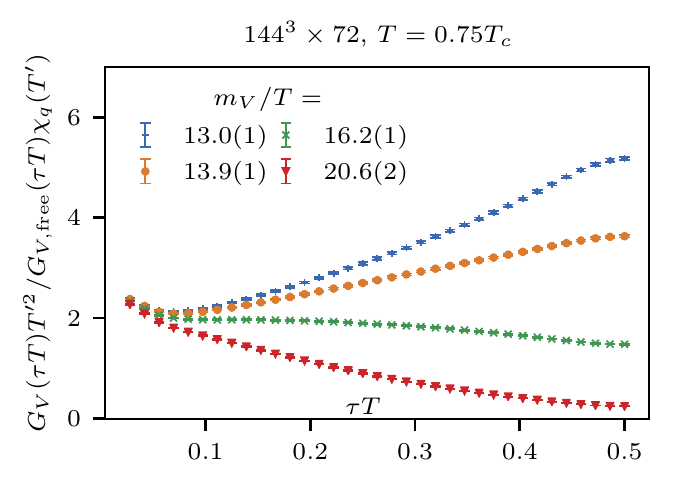}
    \includegraphics{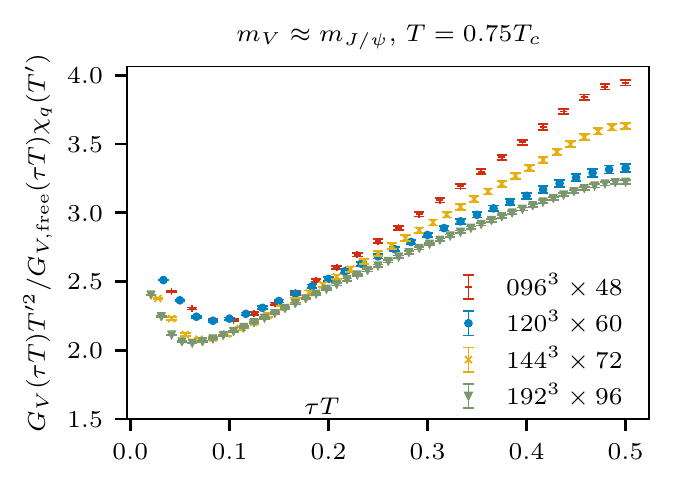}
    \caption{Left: Normalized correlators for different $\kappa$-values as input at a fixed lattice size of $144^3\times72$ at $T=0.75T_c$
         The labels show the corresponding measured ground state screening mass.
        Right: Normalized charmonium correlators directly from the lattice for different lattice size for $T=0.75T_c$.
        In both plots, the lattice correlators have been normalized with the free correlator defined in equation \ref{eq:Gfree}
        with a quark mass $m_q=1.5$GeV and with the quark number susceptibility $\chi_q$ at $T'=2.25T_c$.
    }
    \label{fig:pure_corr}
\end{figure}

\section{Interpolating to the same vector meson mass}
First, let us have a look at the correlators from the lattice for temperatures $T=0.75T_c$, as they are shown in 
figure \ref{fig:pure_corr}. In the left plot we show the correlators for the $N_\sigma=144$ lattice for four different $\kappa$-
values closest to charmonium. We find a very strong mass dependence: Even for small differences in the corresponding
screening mass, we get a rather large deviation of the correlators. 

In the right plot we show the correlators whose measured screening mass lies closest to the $J/\psi$ 
mass. As it can be seen, the ordering of the correlators does not
match to the ordering of the lattice sizes. This makes it impossible to perform a continuum extrapolation
based directly on the lattice correlators. As this effect stems from the slight uncertainties 
of the quark mass tuning, we have to make sure that all correlators are matched to exactly the same quark mass.

\begin{figure}
    \centering
    \includegraphics[width=0.49\textwidth]{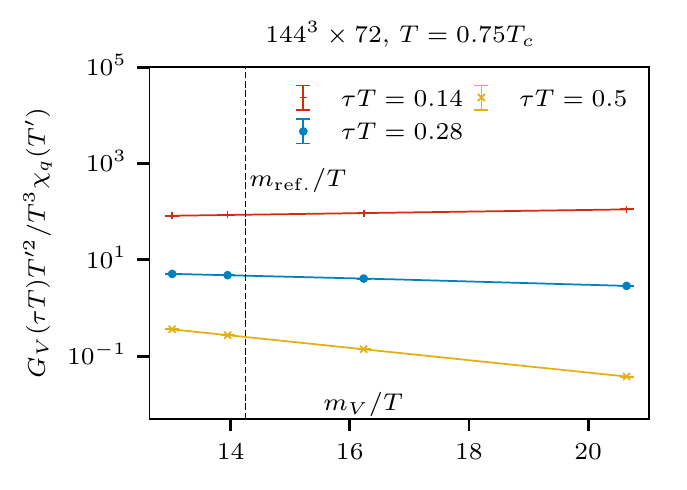}
    \includegraphics[width=0.49\textwidth]{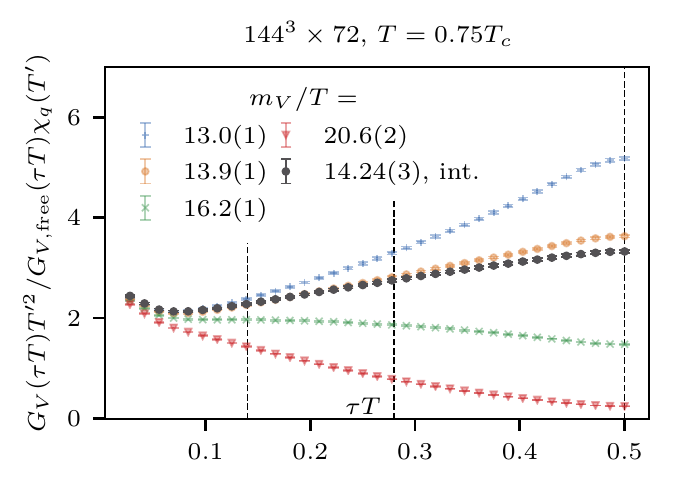}
    \caption{Left: Three examples for the meson mass interpolation between different correlators. The
        figure shows the interpolation for one out of 1000 bootstrap samples. The correlators are normalized
    with the quark number susceptibility at $T'=2.25T_c$. The correlators have been interpolated by fitting equation
    \ref{eq:mass_int} to the data. The vertical line shows the reference mass of charmonium at $T=0.75T_c$ taken
    from the finest lattice.
    Right: Same as left plot in figure \ref{fig:pure_corr} with the interpolated correlator for the charmonium mass. The vertical lines
mark the positions $\tau T$ of the example plots on the left.}
    \label{fig:mass_interpolation}
\end{figure}

This is realized by interpolating between the correlators corresponding to different screening masses. 
To do so, we plot the correlators at each $\tau T$ separately against the measured screening mass
from the $T=0.75T_c$ lattice (see right plot of figure \ref{fig:mass_interpolation}).
Note that this means that also for higher temperatures we use the screening mass
from the lowest temperature lattice. 

The actual interpolation is done by fitting the data with an exponential of a quadratic function:
\begin{equation}
    G\left(\frac{m}{T}, \tau T\right) = \exp\left(\alpha_1\left(\tau T\right)\left(\frac{m}{T}\right)^2+\alpha_2\left(\tau T\right)\left(\frac{m}{T}\right)+\alpha_3\left(\tau T\right)\right)
    \label{eq:mass_int}
\end{equation}
Then, the new interpolated correlators are computed by inserting the same reference mass $m_\mathrm{ref}/T$
for each $\tau T$ and each lattice size into the above equation (vertical line in the right plot of figure \ref{fig:mass_interpolation}).
As also the temperature differs slightly between the lattices, we choose $m/T$ from the finest lattice as the reference mass
for bottomonium and charmonium..
In particular this gives masses of $m_\mathrm{ref, charm}/T=14.25$ for charmonium and
$m_\mathrm{ref, bottom}/T = 44.06$ for bottomonium.

Because of the large differences between the correlators of different mass, we did not use all six available different masses.
Instead we only used the closest four correlators for charmonium and the closest three for bottomonium.

In principle this method can be performed for each configuration individually. However, due to fluctuations
between the individual measurements, this gives larger errors. Instead we used bootstrap samples with
a size of 1000 samples for the mass interpolation. When the choice of the samples is the same for each $\tau T$, the bootstrap 
method preserves the correlations within the correlator.

With this method we are now able to calculate the correlator for any arbitrary meson mass between charmonium and
bottomonium and can retune the correlators for all lattice sizes after the actual computation of the correlator and, therefore,
can compensate uncertainties that stem from slight errors in the tuning of the quark masses.

\begin{figure}
    \centering
    \includegraphics[width=0.49\textwidth]{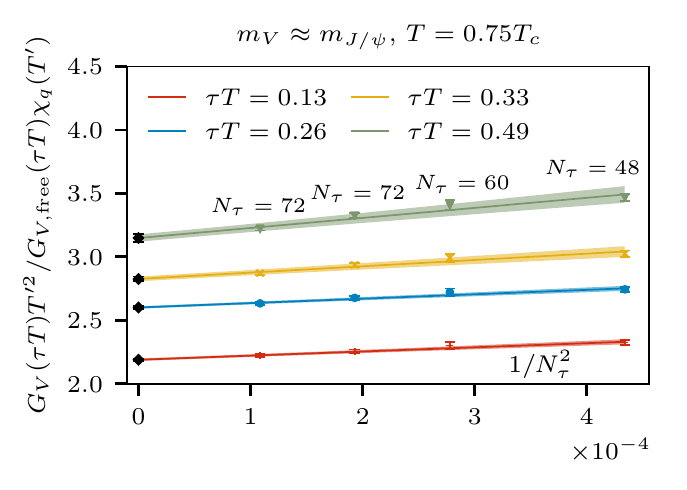}
    \includegraphics[width=0.49\textwidth]{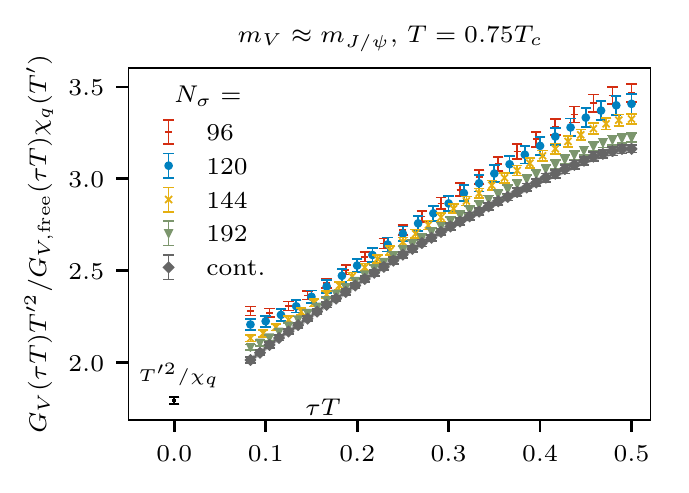}
    \caption{Left: Examples for the extrapolation to the continuum linear in $1/{N_\tau}^2$ for one out of 1000 bootstrap samples.
        For the three coarsest
    lattices, the data points arise from a B-spline interpolations to match to the data points of the finest lattice at each
    $\tau T$.
    Right: The mass interpolated lattice correlators and the corresponding continuum extrapolation. We also show the data point
    for $T'^2/\chi_{q, \mathrm{cont.}}$, at zero distance.
        In both plots, the correlators have been normalized with the free correlator defined in equation \ref{eq:Gfree}
        with a quark mass $m_q=1.5$GeV and with the quark number susceptibility $\chi_q$ at $T'=2.25T_c$.
}
    \label{fig:extrapolation}
\end{figure}
\section{Extrapolation to the continuum}

Having performed the mass interpolation on the three coarsest lattices (as we used $m_\mathrm{ref}/T$ from the finest lattice,
we do not need to interpolate the finest lattice), we end up with well ordered correlators (right plot in figure \ref{fig:extrapolation})
and we are ready to perform the continuum extrapolation.
This is realized using a linear extrapolation in $1/{N_\tau}^2$ on each bootstrap sample
(see left plot of figure \ref{fig:extrapolation}). The three coarsest lattices have
been B-spline interpolated in order to estimate all correlators at distances available on the finest
lattice. Due to the cut off effects at short distances, we restricted the extrapolation
to values larger than $\tau T = 0.08$.


\begin{figure}
    \centering
    \includegraphics[width=0.49\textwidth]{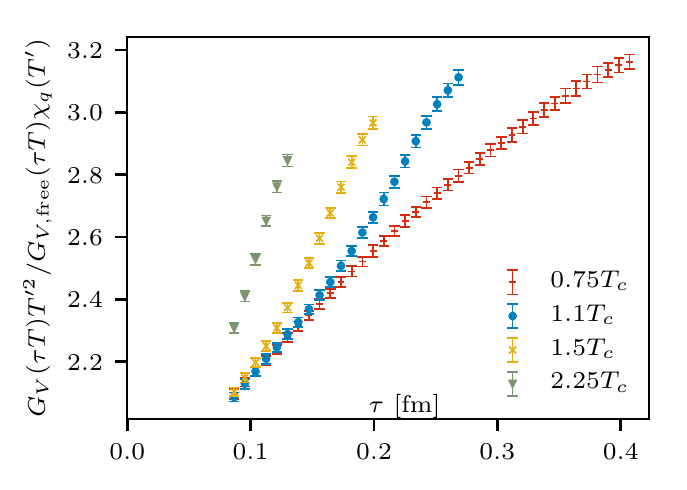}
    \includegraphics[width=0.49\textwidth]{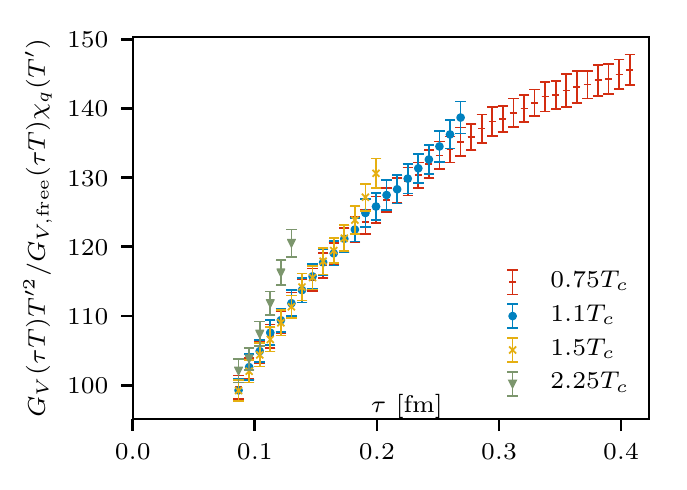}
    \caption{
        Continuum extrapolated vector correlators for different temperatures for charmonium (left) and
        bottomonium (right). 
        The correlators have been normalized with the free correlator defined in equation \ref{eq:Gfree}
        with a quark mass $m_q=1.5$GeV for charmonium and $m_q=5$GeV for bottomonium
        and with the corresponding quark number susceptibility $\chi_q$ at $T'=2.25T_c$.
}
    \label{fig:final_cont}
\end{figure}

The final results for all temperatures can be found in figure \ref{fig:final_cont}. We find that the
correlators agree for short distances of $\tau T$ at almost all temperatures. Only at $T=2.25T_c$ for charmonium
we find deviations. Nevertheless, for larger distances we see clear temperature effects for charmonium. 
On the contrary, for bottomonium we find temperature effects only for $T=2.25T_c$, while for lower temperatures,
the bound state seems to be independent of the temperature.

\section{Conclusion}
For the first time, we were able to perform a continuum extrapolation of temporal Euclidean vector meson
correlation functions for charmonium and bottomonium in quenched lattice QCD. To do so, we interpolated the
correlators to exactly the same vector meson mass and then performed a continuum extrapolation linear in $1/N_\tau^2$. 
Within the temperature range $T=0.75T_c-2.25T_c$, we observe strong temperature effects for charmonium,
but only slight effects for bottomonium at $T=2.25T_c$.

The continuum extrapolated correlators are now ready for further investigations: Using qualitative 
motivated fit Ansätze one can find which parts of the spectral function
contribute to the temperature dependence and which not. Furthermore, statistical Bayesian methods,
such as Stochastical Analytical Interference (SAI)\cite{PhysRevE.81.056701,2004cond.mat..3055B}, the Stochastic Optimization Method (SOM)\cite{diagrammatic} or the Maximum Entropy
Method (MEM)\cite{Asakawa:2000tr} can be used to extract details of the spectral function. These methods have already been
tested with the lattice data of the finest lattice \cite{som} and will be applied to the actual continuum 
correlators.

\section{Acknowledgement}

This work was supported by the BMBF through grant 05P15PBCAA, by the
DAAD through grant 56268409, by the DFG through grant CRC-TR 211 and by the NSFC under grants
11535012 and 11521064.

Computations have been performed using the JARA-HPC resources at RWTH Aachen and JSC Jülich (projects
JARA0039 and JARA0108), the OCuLUS Cluster at the Paderborn Center for Parallel Computing
and the Bielefeld GPU cluster.
\bibliography{lattice2017}

\begin{thebibliography}{26}

\bibitem{Matsui:1986dk}
T.~Matsui, H.~Satz, Phys. Lett. \textbf{B178}, 416 (1986)

\bibitem{Arnaldi:2009ph}
R.~Arnaldi (NA60), Nucl. Phys. \textbf{A830}, 345C (2009), \texttt{0907.5004}

\bibitem{Adare:2008qa}
A.~Adare et~al. (PHENIX), Phys. Rev. Lett. \textbf{101}, 232301 (2008),
  \texttt{0801.4020}

\bibitem{Abelev:2012rv}
B.~Abelev et~al. (ALICE), Phys. Rev. Lett. \textbf{109}, 072301 (2012),
  \texttt{1202.1383}

\bibitem{Aad:2010aa}
G.~Aad et~al. (ATLAS), Phys. Lett. \textbf{B697}, 294 (2011),
  \texttt{1012.5419}

\bibitem{Chatrchyan:2012np}
S.~Chatrchyan et~al. (CMS), JHEP \textbf{05}, 063 (2012), \texttt{1201.5069}

\bibitem{Chatrchyan:2012lxa}
S.~Chatrchyan et~al. (CMS), Phys. Rev. Lett. \textbf{109}, 222301 (2012),
  \texttt{1208.2826}

\bibitem{Abelev:2013lca}
B.~Abelev et~al. (ALICE), Phys. Rev. Lett. \textbf{111}, 102301 (2013),
  \texttt{1305.2707}

\bibitem{Adare:2014rly}
A.~Adare et~al. (PHENIX), Phys. Rev. \textbf{C91}, 044907 (2015),
  \texttt{1405.3301}

\bibitem{Vertesi:2014tfa}
R.~Vértesi (STAR), Nucl. Part. Phys. Proc. \textbf{273-275}, 1588 (2016),
  \texttt{1410.3959}

\bibitem{Moore:2004tg}
G.D. Moore, D.~Teaney, Phys. Rev. \textbf{C71}, 064904 (2005),
  \texttt{hep-ph/0412346}

\bibitem{CaronHuot:2008uh}
S.~Caron-Huot, G.D. Moore, JHEP \textbf{02}, 081 (2008), \texttt{0801.2173}

\bibitem{Ding:2012iy}
H.T. Ding, A.~Francis, O.~Kaczmarek, F.~Karsch, H.~Satz, W.~Söldner, EPJ Web
  Conf. \textbf{70}, 00061 (2014), \texttt{1210.0292}

\bibitem{Ding:2012sp}
H.T. Ding, A.~Francis, O.~Kaczmarek, F.~Karsch, H.~Satz, W.~Soeldner, Phys.
  Rev. \textbf{D86}, 014509 (2012), \texttt{1204.4945}

\bibitem{Ohno:2014uga}
H.~Ohno, H.T. Ding, O.~Kaczmarek, PoS \textbf{LATTICE2014}, 219 (2014),
  \texttt{1412.6594}

\bibitem{Francis:2015daa}
A.~Francis, O.~Kaczmarek, M.~Laine, T.~Neuhaus, H.~Ohno, Phys. Rev.
  \textbf{D92}, 116003 (2015), \texttt{1508.04543}

\bibitem{Ding:2016hua}
H.T. Ding, O.~Kaczmarek, F.~Meyer, Phys. Rev. \textbf{D94}, 034504 (2016),
  \texttt{1604.06712}

\bibitem{Ghiglieri:2016tvj}
J.~Ghiglieri, O.~Kaczmarek, M.~Laine, F.~Meyer, Phys. Rev. \textbf{D94}, 016005
  (2016), \texttt{1604.07544}

\bibitem{Burnier:2017bod}
Y.~Burnier, H.T. Ding, O.~Kaczmarek, A.L. Kruse, M.~Laine, H.~Ohno,
  H.~Sandmeyer (2017), \texttt{1709.07612}

\bibitem{Francis:2015lha}
A.~Francis, O.~Kaczmarek, M.~Laine, T.~Neuhaus, H.~Ohno, Phys. Rev.
  \textbf{D91}, 096002 (2015), \texttt{1503.05652}

\bibitem{Sheikholeslami:1985ij}
B.~Sheikholeslami, R.~Wohlert, Nucl. Phys. \textbf{B259}, 572 (1985)

\bibitem{PhysRevE.81.056701}
S.~Fuchs, T.~Pruschke, M.~Jarrell, Phys. Rev. E \textbf{81}, 056701 (2010)

\bibitem{2004cond.mat..3055B}
K.S.D. {Beach}, eprint arXiv  (2004), \texttt{cond-mat/0403055}

\bibitem{diagrammatic}
A.~Mishchenko, N.~Prokof'ev, A.~Sakamoto, B.~Svistunov, Phys. Rev. B
  \textbf{62} (2000)

\bibitem{Asakawa:2000tr}
M.~Asakawa, T.~Hatsuda, Y.~Nakahara, Prog. Part. Nucl. Phys. \textbf{46}, 459
  (2001), \texttt{hep-lat/0011040}

\bibitem{som}
H.~Ohno, H.T. Ding, H.T. Shu, O.~Kaczmarek, in preperation  (2017)

\end{thebibliography}

\end{document}